\documentclass[preprint]{ptptex}

\usepackage{epsfig}

\newcommand{\nn}{\nonumber}

\title{
Single Longitudinal-Spin Asymmetries in Lepton-Pair Production
 at RHIC and J-PARC}

\author{
Hiroshi \textsc{Yokoya}\footnote{E-mail address:
           \texttt{yokoya@nt.sc.niigata-u.ac.jp}}}

\inst{
Department of Physics, Niigata University,\\
Niigata 950-2181, JAPAN}

\abst{
We study the single longitudinal-spin asymmetries in lepton-pair
 production with large transverse-momentum at RHIC and J-PARC
 experiments.
The asymmetries in the azimuthal angular distribution of a lepton can
 arise from an absorptive part of production amplitudes.
We revisit the one-loop calculation for the absorptive part of
 production amplitudes in perturbative QCD, and show that the
 asymmetries can be sizable at RHIC and J-PARC.
Measurement of the asymmetries would test the one-loop prediction
 for the scattering phase of this process, and provide support for a
 study of the single transverse-spin asymmetries in the same kinematical
 region.
}

\begin{document}

\maketitle

% Introduction
Single longitudinal-spin asymmetries (SLSAs) in the Drell-Yan process
 can arise in the angular distribution of a lepton, when the lepton pair
 has non-zero transverse momentum ($q_T$).
The asymmetries are na\"ively $T$-odd, and thus result from the complex
 phase of the production amplitude.
In the QCD collinear formalism at leading-twist level, a leading-order
 contribution to the complex phase comes from one-loop amplitudes with
 on-shell intermediate state.
The one-loop calculation of the imaginary part of the Drell-Yan
 amplitudes has been performed in Refs.~\citen{pire}
 and \citen{carlitz}, where possibly large asymmetries are predicted.
Related calculations have been also done for na\"ive-$T$-odd asymmetries
 in $W$-jet~\cite{hhk} and $Z$-jet~\cite{hky} events at hadron colliders.
Because na\"ive-$T$-odd asymmetries in hard processes have not been
 tested yet experimentally, measurement of such SLSAs is of great
 interest.\\ 

On the other hand, single transverse-spin asymmetries (STSAs) have been
 part of recent primary progresses in hadron-spin physics,
 since the discovery of the large STSAs in several
 experiments.
STSAs require chirality flip of the amplitude, in addition to the
 scattering phase.
In the collinear formalism, the STSAs call for higher-twist
 distributions to achieve the chirality flip,
 and pole of propagators to the scattering phase~\cite{qs,ji,ekt}.

For the small-$q_T$ case, there exists another approach to
 describe the STSAs, based on the formalism with
 transverse-momentum-dependent parton
 distributions~\cite{ji,sivers,jmy}.
Although an overlap region to be described simultaneously by the two
 formalism is pointed out~\cite{ji}, it is meaningful to know which
 approach is effective at a certain kinematical point.
In most of the experiments, the SLSAs and the STSAs can be measured at
 same kinematical points in one experiment.
Therefore, measurement of SLSAs which have less ambiguity theoretically
 may become a key reference to that of STSAs where the unknown
 non-perturbative functions are involved.\\

In this paper, we revisit the results of Refs.~\citen{pire} and
 \citen{carlitz} and
 give phenomenological studies for the SLSAs in the Drell-Yan process
 for the RHIC and J-PARC experiments.
We consider productions of a lepton pair with large transverse-momentum,
 in collisions of longitudinally-polarized proton and unpolarized
 proton;
\begin{align}
\vec{p} + p \to \ell^{-} + \ell^{+} + X.
\end{align}
%

% Formalism
Defining the spin-dependent cross section as
 $\Delta\sigma=(\overrightarrow{\sigma}-\overleftarrow{\sigma})/2$,
 where $\overrightarrow{\sigma}$ ($\overleftarrow{\sigma}$) denotes the
 cross section for the collision of polarized proton with positive
 (negative) helicity, the SLSAs are expressed in the lepton
 angular distributions as~\cite{pire,carlitz}
\begin{align}
\left(\frac{d\sigma}{dQ^2dq_T^2dy}\right)^{-1}
\frac{d\Delta\sigma}
 {dQ^2dq_T^2dyd\Omega} = \frac{3}{16\pi}
\left[A_{L1}\sin2\theta\sin\phi
+ A_{L2}\sin^2\theta\sin2\phi\right] . \label{dsig}
\end{align}
$Q^2$ is the invariant mass of the lepton pair, $q_T$ is transverse
 momentum of the lepton pair with respect to the collision axis, and $y$
 is the rapidity of the lepton pair in the center of mass frame of
 the protons where the $z$-axis is along the three momenta of polarized
 proton.
The spin-independent cross section is defined as
 $\sigma=(\overrightarrow{\sigma}+\overleftarrow{\sigma})/2$, and
$d\Omega=d\cos\theta d\phi$ where $\theta$, $\phi$ are polar,  azimuthal
 angles of a lepton $\ell^{-}$, respectively, in a rest frame of the
 lepton pair.
Coordinates of the rest frame are fixed to the Collins-Soper
 frame~\cite{cs}, so the $z$ axis is taken to bisect the opening angle
 between $\vec{p}_{\vec{p}}$ and $-\vec{p}_{p}$, and the $y$ axis is
 along the direction of $\vec{p}_{\vec{p}}\times (-\vec{p}_{p})$.
The azimuthal angle is measured from the $x$ axis which lies in the
 scattering plane.

The structure functions for the SLSAs, $A_{L1,L2}$, are calculated as,
\begin{align}
&A_{L1,L2}(Q^2,q_T^2,y) = 
\frac{\sum_{a,b}\int dY \Delta D_{a/p}(x_+,\mu^2)D_{b/p}(x_-,\mu^2)
g^{ab}_{1,2}(z,\cos\hat\theta)}
{\sum_{a,b}\int dY D_{a/p}(x_+,\mu^2)D_{b/p}(x_-,\mu^2)
f^{ab}(z,\cos\hat\theta)},\label{asym}
\end{align}
where $D_{a/p}$ and $\Delta D_{a/p}$ are the unpolarized and
longitudinally-polarized parton distribution functions (PDFs) of proton,
respectively,
\begin{align}
 &x_{\pm}=\sqrt{\frac{\hat{s}}{s}}\, e^{\pm Y},
\hspace{10pt}\hat{s}=Q^2+\frac{2q_T^2}{\sin^2\hat\theta}
\left(1+\sqrt{1+\frac{Q^2\sin^2\hat\theta}{q_T^2}}\right),
\hspace{10pt}\cos\hat\theta=\tanh{(y-Y)},
\end{align}
and $z=Q^2/\hat{s}$.
$A_{L1,L2}$ are expressed in terms of weighted integrals of
spin-dependent cross section as, 
\begin{align}
A_{L1,L2}(Q^2,q_T^2,y) =
\Bigg(\int d\Omega \,\omega_{1,2}(\theta,\phi)\,
 \frac{d\Delta\sigma}{dQ^2dq_T^2dy d\Omega}\Bigg)\Bigg/
 \Bigg(\frac{d\sigma}{dQ^2dq_T^2dy}\Bigg),
\end{align}
where the weight functions are $\omega_{1}=5\sin 2\theta \sin\phi$ and
 $\omega_{2}=5\sin^2\theta\sin2\phi$.\\

In the leading order (LO), the hard part functions $f^{ab}$ and
 $g^{ab}_{i}$ in Eq.~(\ref{asym}) have contributions from the
 annihilation subprocess $q\bar{q}\to \ell^{-}\ell^{+}g$ and the Compton
 subprocess $qg\to\ell^{-}\ell^{+}q$ ($\bar{q}g\to \ell^{-}\ell^{+}
 \bar{q}$).
$f^{ab}$ are calculated from tree-level diagrams, and $g^{ab}_{i}$ from
 the one-loop diagrams, in the LO.

Here, we reproduce the results of Refs.~\citen{pire}
 and \citen{carlitz}, following the notation of Refs.~\citen{hhk}
 and \citen{hky}.
For the spin-independent functions,
\begin{align}
 f^{q\bar{q}}(z,\cos\hat\theta)=e_q^2 \frac{C_F}{N}
\frac{1}{1+1/z}\,f^{q\bar{q}}_{1}\,,\quad
 f^{qg}(z,\cos\hat\theta)=e_q^2\frac{T_F}{N}
\frac{1}{1+1/z}\,f^{qg}_{1}\,,
\end{align}
with
\begin{align}
f^{q\bar{q}}_{1}=\frac{a^2+b^2}{2ab(1-c)},\quad
f^{qg}_{1}=\frac{b^2+(a+b-2ab)^2}{2(1-a)bc},
\end{align}
and $f_1^{\bar{q}q}=f_1^{q\bar{q}} (a\leftrightarrow b)$,
$f_1^{gq}=f_1^{qg} (a\leftrightarrow b)$, where
\begin{align}
 a=\frac{2z}{1+z-(1-z)\cos\hat\theta}, \quad
 b=\frac{2z}{1+z+(1-z)\cos\hat\theta},
\end{align}
and $c=a+b-ab$. 
$e_q$ is the electromagnetic charge of quarks, and
color factors are $N=3$, $C_F=4/3$, $T_F=1/2$ and $C_1=-1/6$.

For the spin-dependent functions,
\begin{align}
g^{q\bar{q}}_{1,2}(z,\cos\hat\theta)
=-e_q^2\alpha_s\frac{C_F}{N}
 \frac{1}{1+1/z}\,f^{q\bar{q}}_{8,9}\,,\quad
g^{qg}_{1,2}(z,\cos\hat\theta)
=-e_q^2\alpha_s\frac{T_F}{N}
\frac{1}{1+1/z}\,f^{qg}_{8,9}\,,
\end{align}
where
\begin{align}
f^{q\bar{q}}_{8}&=\frac{c}{2\sqrt{1-c}}
\bigg[-C_F\frac{a}{b}+C_1\frac{1}{1-a}\ln{\frac{a}{c}}\bigg]
 - (a\leftrightarrow b),\\
f^{q\bar{q}}_{9}&=\frac{\sqrt{c}}{2}
\bigg[-C_F\frac{a}{2b}-C_1\frac{1}{1-a}\left(1+\frac{c}{c-a}
\ln{\frac{a}{c}}\right)\bigg] + (a \leftrightarrow b),\\
f^{qg}_{8}&=\frac{c-b}{2\sqrt{1-c}}
\bigg[-C_F\left(\frac{a}{b}-\frac{1+a}{2}\right)
+C_1\left\{b-1+\frac{a}{c}
\left(b+\frac{c-b}{c}\ln{\frac{1}{1-c}}\right)\right\}\bigg],\\
f^{qg}_{9}&=\frac{c-b}{2\sqrt{c}}\bigg[-C_F\left(\frac{a}{2b}
+\frac{1+a}{2}\right)\nn\\
&\hspace{100pt}+C_1\left\{b+\frac{a}{c}\ln{\frac{1}{1-c}}
-\frac{1}{1-a}\left(1+\frac{a}{c-a}\ln{\frac{a}{c}}\right)\right\}\bigg].
\end{align}
Similarly for other subprocesses,
 $f^{\bar{q}q}_{8}=-f^{q\bar{q}}_{8}(a\leftrightarrow b)$,
 $f^{\bar{q}q}_{9}=f^{q\bar{q}}_{9}(a\leftrightarrow b)$, and
\begin{align}
f^{gq}_{8}&= f^{qg}_{8}(a\leftrightarrow b)
- C_1\frac{a(1-a)}{\sqrt{1-c}},\\
f^{gq}_{9}&=- f^{qg}_{9}(a\leftrightarrow b)
- C_1\frac{c-a}{\sqrt{c}}\frac{1}{1-b}
\left(1+\frac{b}{c-b}\ln{\frac{b}{c}}\right).
\end{align}
The functions for anti-quark and gluon scattering are
 $f^{\bar{q}g}_{i}=f^{qg}_{i}$, $f^{g\bar{q}}_{i}=f^{gq}_{i}$ for
 $i=1,8,9$.\\

\begin{figure}[t]
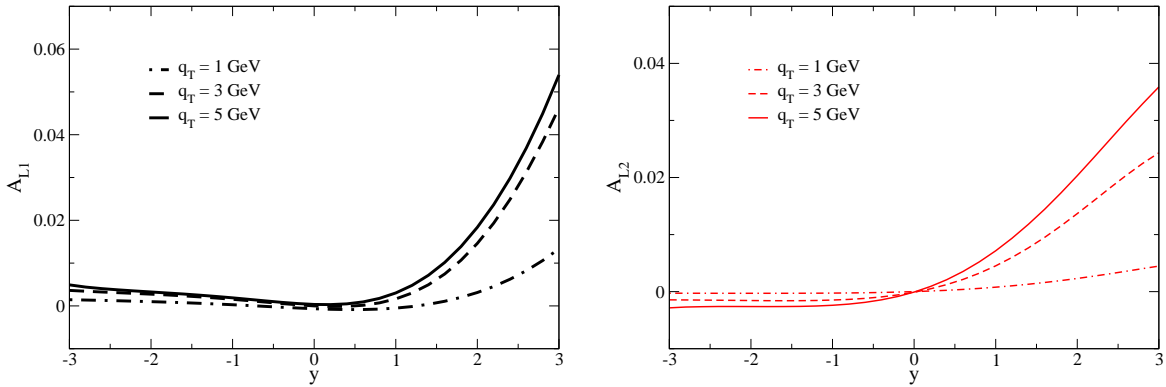

\vspace{10pt}
\begin{center}
\epsfig{file=al1-r.eps,width=.45\textwidth} \quad
\epsfig{file=al2-r.eps,width=.45\textwidth}
\caption{Single longitudinal-spin asymmetries, $A_{L1}$ (left) and
 $A_{L2}$ (right) for $\sqrt{s}=200$ GeV and $Q=5$ GeV. Asymmetries
 for three different values of $q_T$ are plotted; $q_T=1$ GeV (dot-dashed),
 $q_T=3$ GeV (dashed) and $q_T=5$ GeV (solid).
 $y$ denotes the rapidity of the lepton pair in the center of mass frame
 of protons.
}\label{fig:rhic}
\end{center}
\end{figure}

% Numerical estimate
For a numerical estimate, we use the GRV98 (NLO $\overline{\rm MS}$
 scheme) parameterization~\cite{grv} for the unpolarized PDFs, and the
 AAC03 parameterization~\cite{aac} for polarized PDFs.
We set the scale of PDFs and the strong coupling constant $\alpha_s$ to
 $\mu=Q$.
In Fig.~\ref{fig:rhic}, we show the numerical estimates of these
 asymmetries for RHIC $\vec{p}p$ collisions at $\sqrt{s}=200$ GeV and
 $Q=5$ GeV.
We plot the asymmetries $A_{L1}$ (left) and $A_{L2}$ (right) for
 three different values of $q_T$; 1 GeV (dot-dashed), 3 GeV (dashed) and
 5 GeV (solid).
The absolute magnitude of the asymmetries increases with $q_T$, and
 becomes largest in large rapidity region, $A_{L1}\sim 5\%$ and
 $A_{L2}\sim 3.5\%$.
For small $q_T$, the asymmetries are predicted to be small over the
 whole $y$ range.
In the forward region (positive $y$), the asymmetries mainly come from
 subprocesses with polarized quarks.
On the other hand, in the backward region (negative $y$) the
 asymmetries receive dominant contributions from subprocesses with a
 polarized gluon.\\

\begin{figure}[t]
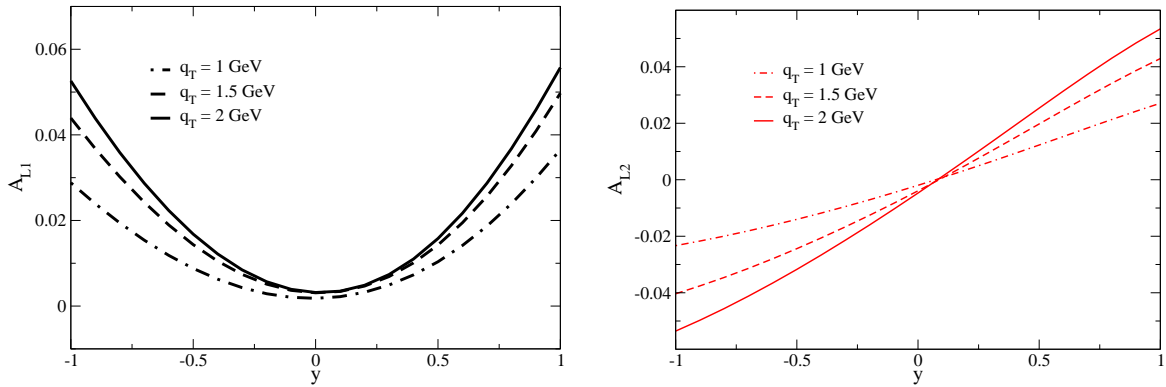

\vspace{10pt}
\begin{center}
\epsfig{file=al1-j.eps,width=.45\textwidth} \quad
\epsfig{file=al2-j.eps,width=.45\textwidth}
\caption{Same as Fig.~\ref{fig:rhic} but for $\sqrt{s}=10$ GeV and
 $Q=2$ GeV.
 Asymmetries for three different values of $q_T$ are plotted; $q_T=1$
 GeV (dot-dashed), $q_T=1.5$ GeV (dashed) and $q_T=2$ GeV (solid).
}\label{fig:jparc}
\end{center}
\end{figure}

In Fig.~\ref{fig:jparc}, we show the numerical estimates of these
 asymmetries for J-PARC experiment with polarized-proton at
 $\sqrt{s}=10$ GeV and $Q=2$ GeV. 
The asymmetries for $q_T=1$ GeV (dot-dashed), 1.5 GeV (dashed) and 2 GeV
 (solid) are plotted.
The asymmetries amount to $A_{L1}\sim 5.5\%$ and
 $A_{L2}\sim \pm 5\%$ in the large $|y|$ region for $q_T=2$ GeV.
For the J-PARC case, the absolute magnitude of the asymmetries in
 the backward region are almost the same as those in the forward
 region.
This is because the asymmetries are proportional to $\Delta g/g$ in the
 backward region, and the mean values of $x_+$ are around
 $\langle x_+ \rangle_{y=-1} \sim 0.3$-0.5 at J-PARC, depending on
 $q_T$, while $\langle x_+ \rangle_{y=-3} \sim 0.01$-0.03 at RHIC.
The asymmetries may be used to constrain the parameterizations of
 polarized PDFs.
In the backward lepton-pair production, $\Delta{g}/g$ at around 
 $\langle x\rangle\sim 0.01$ (0.3) is tested at RHIC (J-PARC). 
\\

% Remarks
Finally, we make some remarks about the estimated asymmetries.
Since our predictions are based on the LO calculation, there exists
 a significant ambiguity in the choice of the scale of the PDFs and
 $\alpha_s$.
The asymmetries are ${\cal O}(\alpha_s)$, and therefore increase with
 decreasing scale of $\alpha_s$. 
In case we take the scale as $\mu=q_T$, the asymmetries increase
 at most by 30\%-50\%, uniformly in $y$.
QCD higher-order corrections have been known to change the $q_T$
 distribution of production cross-sections, especially for small-$q_T$
 region ($q_T\ll Q$) by the multiple soft-gluon emissions~\cite{kkt}.
However, even though we do not proceed to the small-$q_T$ region, it is
 expected that these effects largely cancel out in the lepton's angular
 asymmetries, as far as the QCD collinear formalism is valid~\cite{bw}.

For the production of lepton pairs with small $q_T$, the asymmetries are
 described by a formalism with transverse-momentum-dependent parton
 distributions~\cite{boer}.
In this formalism, only the asymmetry $A_{L2}$ is generated, while
 $A_{L1}$ remains zero\footnote{
The asymmetry $A_{L2}$ is proportional to a product of chiral-odd
 distributions $h^{\perp}_1$ and $h^{\perp}_{1L}$, where the former is
 $T$-odd (Boer-Mulders function). 
See Ref.~\citen{boer} for the definition of these functions.
}.
\\

% Conclusion
In conclusion, we have studied the single longitudinal-spin asymmetries
 in lepton-pair production at RHIC and J-PARC.
The asymmetries in the azimuthal angular distribution of a lepton arise
 from the absorptive part of the production amplitude, when the lepton
 pair has non-zero transverse momentum.
We re-analyzed the asymmetries in leading-order for the RHIC and J-PARC
 experiments, and showed that they can be sizable for large $q_T$ and at
 large forward or backward rapidity.
This would be a good experimental test for the scattering phase
 of the production amplitude, and the comparison with the one-loop
 calculation in the collinear formalism may provide phenomenological
 supports for a study of the single transverse-spin asymmetries at the
 same kinematical region.
\\

% Acknowledgment
We thank K.~Hagiwara, Y.~Koike for useful comments, and W.~Vogelsang for
 discussions and reading the manuscript.
We thank RIKEN BNL Research Center for helpful hospitality during the
 stay.
The work of H.Y.\ is supported in part by the Japan Society for the
 Promotion of Science.

\end{document}